# Free-Standing Epitaxial $SrTiO_3$ Nanomembranes via Remote Epitaxy using Hybrid Molecular Beam Epitaxy

Hyojin Yoon[1, †], Tristan K. Truttmann[1, †], Fengdeng Liu[1], Bethany E. Matthews[2], Sooho Choo[1], Qun Su[3], Vivek Saraswat[4], Sebastian Manzo[4], Michael S. Arnold[4], Mark E. Bowden[5], Jason K. Kawasaki[4], Steven J. Koester[3], Steven R. Spurgeon[2,6], Scott A. Chambers[7] and Bharat Jalan[1,*]

[1]Department of Chemical Engineering and Materials Science, University of Minnesota – Twin Cities, Minneapolis, MN 55455, USA

[2]Energy and Environment Directorate, Pacific Northwest National Laboratory, Richland, Washington 99352, USA

[3]Department of Electrical and Computer Engineering, University of Minnesota – Twin Cities, Minneapolis, MN 55455, USA

[4]Department of Materials Science and Engineering, University of Wisconsin – Madison, Madison, WI 53706, USA

[5]Environmental Molecular Sciences Laboratory, Pacific Northwest National Laboratory, Richland, Washington 99352, USA

[6]Department of Physics, University of Washington, Seattle, Washington 98195, USA

[7]Physical and Computational Sciences Directorate, Pacific Northwest National Laboratory, Richland, Washington 99352, USA

† HY and TKT are equally contributing authors.

* All correspondence should be addressed to BJ (bjalan@umn.edu).

## ABSTRACT

The epitaxial growth of functional materials using a substrate with a graphene layer is a highly desirable method for improving structural quality and obtaining free-standing epitaxial nano-membranes for scientific study, applications, and economical reuse of substrates. However, the aggressive oxidizing conditions typically employed to grow epitaxial perovskite oxides can damage graphene. Here, we demonstrate a technique based on hybrid molecular beam epitaxy that does not require an independent oxygen source to achieve epitaxial growth of complex oxides without damaging the underlying graphene. The technique produces films with self-regulating cation stoichiometry control and epitaxial orientation to the oxide substrate. Furthermore, the films can be exfoliated and transferred to foreign substrates while leaving the graphene on the original substrate. These results open the door to future studies of previously unattainable free-standing nano-membranes grown in an adsorption-controlled manner by hybrid molecular beam epitaxy, and has potentially important implications for the commercial application of perovskite oxides in flexible electronics.

## INTRODUCTION

In conventional epitaxial growth, the epilayer is intimately linked to the substrate. This prevents the reuse of the expensive single-crystal substrate unless the film is to be sacrificed by polishing it away. Furthermore, permanently joining the film and substrate can generate challenges in characterizing the film due to signals from the substrate that are often many orders of magnitude stronger than those of the film (*1*). Therefore, there is both an economic and scientific impetus to develop facile methods of producing free-standing single-crystal nanomembranes (*2*).

The most rudimentary route is the grinding (*3*) or etching (*4*) of bulk wafers down to microscopic thicknesses. But this approach lacks precision and often achieves poor material utilization, because a majority of the wafer is sacrificed. A more precise technique uses a sacrificial layer between the substrate and film which, upon removal, releases the thin film as a free-standing membrane. The sacrificial layer may be selectively melted (*5*) or etched (*6, 7*) away. This offers the advantage of reuse of expensive substrates and has all the precision of the utilized thin-film growth technique. This approach has enabled new studies that are possible only with free-standing membranes, including straining thin films beyond levels possible with conventional biaxial strain (*8*). In systems where the target film preferentially absorbs light, the film region adjacent to the transparent substrate can be preferentially vaporized with intense laser light, serving itself as the sacrificial layer (*9-12*).

Another strategy for achieving free-standing single-crystalline membranes is to utilize an interface with weak adhesion such as by utilizing van der Waals materials (*13*). The technique takes on different names; when the film is oriented to the van der Waals material, it is called van der Waals epitaxy (*14, 15*). If the film is oriented to the underlying substrate, the technique is called remote epitaxy (*16*). In this approach, it is argued that the film nucleates on top of the van der Waals material and is oriented by the interatomic potential from the substrate penetrating through the van der Waals material. However, the underlying mechanism is still under debate due to the possibility of epitaxial lateral overgrowth through micro/nano holes in the van der Waals material (*17*) (*18*). In practice, distinguishing these two mechanisms is challenging and has been a topic of significant current interest. However, there is no doubt that the use of a van der Waals material is nonetheless a proven method to obtain free-standing single-crystal membranes (*13, 16*) that has the advantages of high material utilization (no sacrificial layer required), optional substrate reuse (*19*), and the ability to obtain thin films with better crystalline quality than in traditional epitaxy (*20, 21*).

Given the chemical flexibility and functional diversity of the perovskite oxide material family, extension of remote epitaxy to perovskite oxides is highly desirable as a method to potentially improve structural quality and study free-standing membranes in isolation. Oxide MBE is a highly modular and adaptable low-energy deposition technique that has been used to grow a wide range of perovskite oxides (*22, 23*). Conventional oxide MBE uses effusion cells to sublime or evaporate metals, and more recently, metal suboxides (*24*) and an oxygen source to oxidize the metals at the growth front, as shown in Fig. 1A. The oxygen source is often activated with an inductively coupled radio frequency (RF) (*25, 26*) or electron cyclotron resonance (ECR) microwave (*27*) plasma, or it may be dilute or distilled ozone (*28*). Since these aggressive oxygen sources can decompose graphene, molecular oxygen supplied at a low background pressure of $7 \times 10^{-7}$ Torr has been used as a more gentle source for remote epitaxy of oxides (*29*). However, it has not been possible to grow nano-membranes of defect-managed complex oxides with self-regulating stoichiometry control. For instance, $SrTiO_3$ films growth using conventional MBE requires precise flux control which is no better than 0.1%, potentially resulting into a defect density as high as $10^{19}$ $cm^{-3}$.

Hybrid MBE is a technique that addresses these problems by replacing the elemental Ti with a titanium tetraisopropoxide (TTIP) metal-organic source. The high vapor pressure of TTIP or its decomposition intermediates provides a desorption mechanism that self-regulates the Sr:Ti cation stoichiometry and provides a growth window within which the incorporated Sr:Ti ratio is unity and is impervious to flux instabilities (*30*). Use of this technique has resulted in mobilities in $SrTiO_3$ films of >120,000 $cm^2V^{-1}s^{-1}$ (*31*) using epitaxial strain, suggesting exciting opportunities for tunable electronic properties using membrane engineering.

In this study, we used hybrid MBE with the oxygen source turned off to avoid graphene damage, as shown in Fig. 1B. The four oxygen atoms in each TTIP molecule provide sufficient oxygen to obtain phase-pure $SrTiO_3$. We show that a growth window is achievable even without the use of additional oxygen, a key feature leading to hybrid MBE's high material quality. Critical to remote epitaxy, this aspect of hybrid MBE avoids graphene oxidation while allowing exfoliation and transfer of the epilayer to remote substrates. Unlike the prior reports of oxide remote epitaxy using dry-transfer graphene (*16, 32*), our approach yielded epitaxial $SrTiO_3$ films on wet-transferred graphene (see Fig. S1).

## RESULTS AND DISCUSSION

Fig. 2 shows the results of applying this technique to homoepitaxial $SrTiO_3$ (without graphene). The clear RHEED oscillations (Fig. 2B) and atomically smooth surfaces visible from atomic force microscopy (AFM, Fig. 2C inset) shows that this technique results in atomic precision even without the use of a dedicated oxygen source. Very importantly, the lattice parameter that is indistinguishable from the substrate (Fig. 2C) and a MBE growth window revealed by high-resolution X-ray diffraction (HR-XRD, Fig 2D) show that this modified technique also achieves adsorption-controlled growth with excellent structural quality and reproducibility.

Fig. 3 shows results from the growth of $SrTiO_3$ on bare $SrTiO_3$ substrates (Fig. 3A) as well as on $SrTiO_3$ substrates coated with monolayer graphene (Fig. 3B), bilayer graphene (Fig. 3C), and on LSAT [$(La_{0.18}Sr_{0.82})(Al_{0.59}Ta_{0.41})O_3$] coated with monolayer graphene (Fig. 3D). The RHEED patterns for films grown on bare substrates and on monolayer graphene all show half-order streaks and Kikuchi lines characteristic of high-quality epitaxial $SrTiO_3$, but the RHEED pattern on bilayer graphene is distinct. We later show, using scanning transmission electron microscopy (STEM), that this pattern likely comes from the in-plane rotation of $SrTiO_3$ film owing to poor graphene quality underneath.

The results from Fig. 3 raise the question of whether this modified hybrid MBE technique left the graphene intact, or whether the conditions rapidly decomposed the graphene and simply deposited $SrTiO_3$ directly on top of the remaining oxide substrate. Figs. 4(A-B) address this question with confocal Raman spectroscopy before and after growth on bilayer graphene. Bilayer graphene was chosen because monolayer graphene causes the graphene and $SrTiO_3$ films to crack (see Fig. S2). Although the graphene D peak overlaps with a peak from $SrTiO_3$, the similar positions and intensities of the graphene G and 2D peaks before and after growth indicate that the graphene remains intact and undamaged during growth. Further structural and chemical analysis confirms the retention of the graphene layer after growth, as well as the quality and epitaxial nature of the as-grown film. Fig. 5 shows scanning transmission electron microscopy high-angle annular dark field (STEM-HAADF) images and energy-dispersive X-ray spectroscopy (STEM-EDS) elemental maps of the as-grown film prior to exfoliation and transfer. Specifically, we observe large continuous regions of epitaxial film growth with an apparent [100] (001) // [100] (001) orientation relationship and well-preserved graphene layer, as shown in Figs. 5A-C. We also observe some defective carbonaceous regions, as indicated in Figs. 5D-E, where the graphene layer has "bunched" into faceted piles. This may be residual PMMA or graphene "bunched" into faceted piles. These piles perturb the film growth, leading to the formation of antiphase boundaries at the pile apex and associated in-plane (IP) lattice rotation, likely local strain variations. As shown in Figs. 5F-G, this results in large regions of ordered, on-zone domains interspersed with slightly rotated domains, which are nonetheless epitaxial out-of-plane. Together, these results attest to the compatibility of graphene with the hybrid MBE technique in the absence of additional oxygen. Fig. 4D shows that the film can be exfoliated and transferred to other substrates, as revealed by the presence of the $SrTiO_3$ (002) peak throughout the entire transfer process. Finally, Fig. 4C shows that film exfoliation leaves behind graphene on the substrate. Analysis of the G:2D peak intensity ratios is consistent with the presence of bilayer graphene on the substrate (see Fig. S3). This allows the substrate to be directly reused for growth of more epitaxial membranes.

Finally, Fig. 6 shows STEM-HAADF and STEM-EDS composition maps of the transferred film after annealing. This data again reveals epitaxial, single-phase film on a foreign substrate (*r*-$Al_2O_3$) confirming the successful transfer of an epitaxial $SrTiO_3$ film. This result is consistent with the X-ray diffraction data in Fig. 4D. Future research will investigate the origin of the partial square hole (void) visible in the STEM-HAADF image; it might originate from the growth, exfoliation, transfer, or annealing process.

In summary, we have demonstrated a novel hybrid MBE approach for growing epitaxial films of $SrTiO_3$ on graphene using Sr and TTIP sources without the use of an additional oxygen source. The technique produces films with atomic thickness control and self-regulated cation stoichiometry within a growth window without damaging the underlying graphene. The films can be exfoliated as free-standing membranes and transferred to other substrates. This work opens the door to a wide range of studies on free-standing oxide membranes with the benefit afforded by hybrid MBE. Future studies will explore improving the quality of these films with dry-transferred graphene and investigate a wider variety of material systems and heterostructures. However, the fact that the use of wet-transfer graphene also supports the epitaxial growth clearly suggests the robustness of remote epitaxy process adding to its versatile nature.

## MATERIALS AND METHODS

### Graphene growth and transfer

Graphene was grown on both sides of polycrystalline copper foil using chemical vapor deposition in a quartz tube furnace. First, the foil is hydrogen-annealed in 16 standard cubic centimeters per minute (sccm) of hydrogen at 30 mTorr while the furnace ramps to the growth temperature of 1050 ºC (~20 minutes). Then, the foil is annealed for another 30 minutes at the growth temperature under the same flow and pressure. To grow graphene, 21 sccm hydrogen and 0.105 sccm methane are supplied while the furnace is maintained at 250 mTorr for 30 minutes. These conditions create self-terminating growth of one graphene monolayer. Then, the methane flow is turned off and the hydrogen flow is set to 16 sccm while the furnace cools off (~3 hours).

A solution of 4 wt% polymethylmethacrylate with a molecular weight of 950,000 atomic mass units dissolved in chlorobenzene (PMMA 950 C4, MicroChem Corp.) is used for spin-coating. Then, graphene is removed from the bottom side of the copper foil with a 10-second exposure to oxygen plasma in a reactive ion etcher. Ammonium persulfate solution (7 g $(NH_4)_2S_2O_8$ in 1 L of DI water) was used as a copper etchant to etch away the copper foil. The reader is referred to Fig. S4 for a step-by-step summary of the graphene growth and transfer process.

## $SrTiO_3$ epitaxial film growth

All films were grown on 5 mm × 5 mm substrates of single-crystal $SrTiO_3$ (001) or LSAT (001) with and without a graphene layer. During growth, the substrates were maintained at a thermocouple reading of 900 ºC using a SiC-filament substrate heater. Strontium was supplied by thermal sublimation of distilled Sr dendrites (99.99% pure, Sigma-Aldrich). Titanium and oxygen were supplied by the chemical precursor TTIP (99.999% pure, Sigma-Aldrich) which was fed to a line-of-sight gas injector (E-Science, Inc.) via a custom gas inlet system using a linear leak valve and a Baratron capacitance manometer (MKS Instruments, Inc.) in a PID feedback loop to control the TTIP flow entering the chamber. Immediately after the substrate temperature setpoint reached idle, reflection high-energy electron diffraction (RHEED) was collected in the same chamber where growth took place. The reader is referred to Fig. S5 for a step-by-step summary of the membrane exfoliation and transfer process.

## Characterization

The sample surface topography was measured by a Bruker Nanoscope V Multimode 8 atomic force microscopy (AFM) in contact mode. All X-ray diffraction was performed with a Rigaku SmartLab XE diffractometer. Reciprocal space maps were collected with the HyPix-3000 detector in 1-dimentional mode to simultaneously resolve 2*q* while *w* was scanned. Confocal Raman data was collected with a Witec Alpha 300 R confocal Raman microscope. The 532 nm source light was generated by a frequency-doubled Nd:YAG laser and the output was analyzed with a diffraction grating spectrometer and CCD detector.

Cross-sectional scanning transmission electron microscopy (STEM) samples were prepared using a FEI Helios NanoLab DualBeam $Ga^+$ Focused Ion Beam (FIB) microscope with a standard lift out procedure. The STEM high-angle annular dark field (STEM-HAADF) images shown in Fig. 5 were acquired on a probe-corrected Thermo Fisher Themis Z microscope operating at 300 kV, with a convergence semi-angle of 25.2 mrad and an approximate collection angle range of 65-200 mrad. The STEM energy-dispersive X-ray spectroscopy (STEM-EDS) composition maps shown in Fig. 5 were acquired using a SuperX detector. The STEM-HAADF image shown in Fig. 6 was acquired on a probe-corrected JEOL GrandARM-300F microscope operating at 300 kV, with a convergence semi-angle of 29.7 mrad and a collection angle range of 75–515 mrad. The STEM-EDS composition maps shown in Fig. 6 were acquired using dual JEOL Centurio silicon drift detector setup.

**Acknowledgments**

The authors thank Jeehwan Kim for helpful discussion. Hybrid MBE growth and characterization of $SrTiO_3$ films were supported by the US Department of Energy (DOE) through Grant DE-SC0020211. Graphene growth and integration with the oxide substrate was supported by the Air Force Office of Scientific Research through Grants FA9550-21-0460 and FA9550-21-1-0025 and the National Science Foundation through Grant DMR-1752797. Parts of this work were carried out at the Characterization Facility, University of Minnesota, which receives partial support from the NSF through the MRSEC program under Award DMR-2011401. Substrate preparation was carried out at the Minnesota Nano Center, which is supported by the NSF through the National Nano Coordinated Infrastructure under Award ECCS-2025124. STEM and XPS was supported by the U.S. Department of Energy (DOE), Office of Science, Office of Basic Energy Sciences, Division of Materials Sciences and Engineering under Award #10122. Pacific Northwest National Laboratory (PNNL) is a multiprogram national laboratory operated by Battelle for the U.S. DOE under Contract DE-AC05-79RL01830. We acknowledge facility support from the Environmental Molecular Sciences Laboratory, a DOE Office of Science User Facility sponsored by the Biological and Environmental Research program and located at PNNL. A portion of the microscopy work was performed in the Radiological Microscopy Suite (RMS), located in the Radiochemical Processing Laboratory (RPL) at PNNL.

**Author contributions:** HY, TKT and BJ conceived the idea and designed experiments. HY and TKT developed the growth technique. HY grew and transferred the graphene, grew the films, and transferred the films to foreign substrates. SM transferred some of the graphene to foreign substrates under the supervision of JKK. HY, SC and FL characterized the films with AFM, HR-XRD, and confocal Raman spectroscopy. HY curated the data under the supervision of BJ. BEM performed STEM measurement and analysis under the supervision of SRS and SAC. MEB conducted additional XRD analysis. TKT, HY and BJ wrote the manuscript with input and feedback from all the authors. BJ directed and organized the different aspects of the project.

**Competing interests:** The authors declare no competing interests.

**Data and materials availability:** All data needed to evaluate the conclusions in the paper are present in the paper and/or the Supplementary Materials.

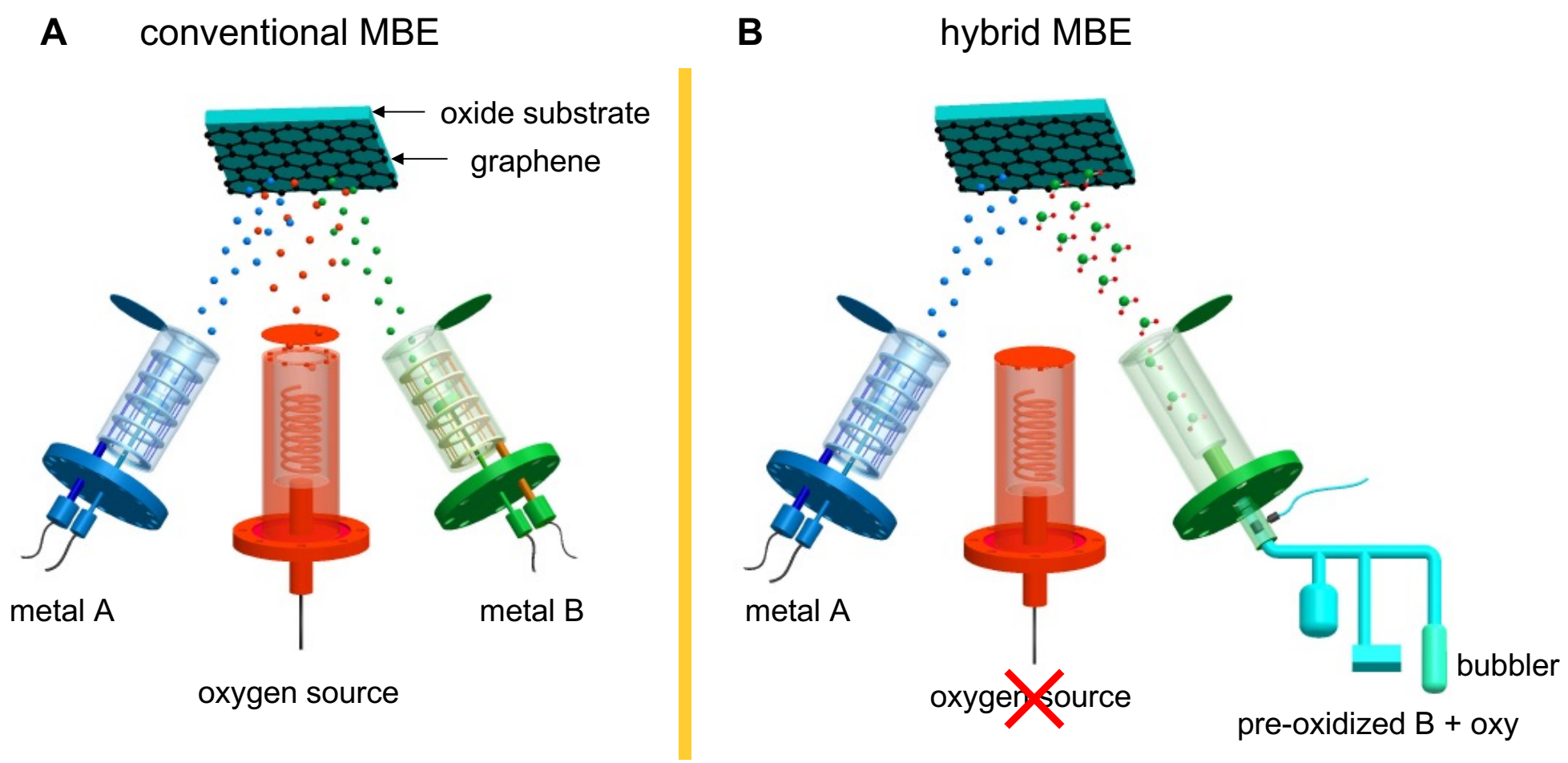

**Fig. 1. Comparison of growth techniques for perovskite oxides on graphene.** (**A**) Conventional oxide MBE. (**B**) Hybrid MBE modified by excluding oxygen. The gentle oxidation environment avoids graphene damage.

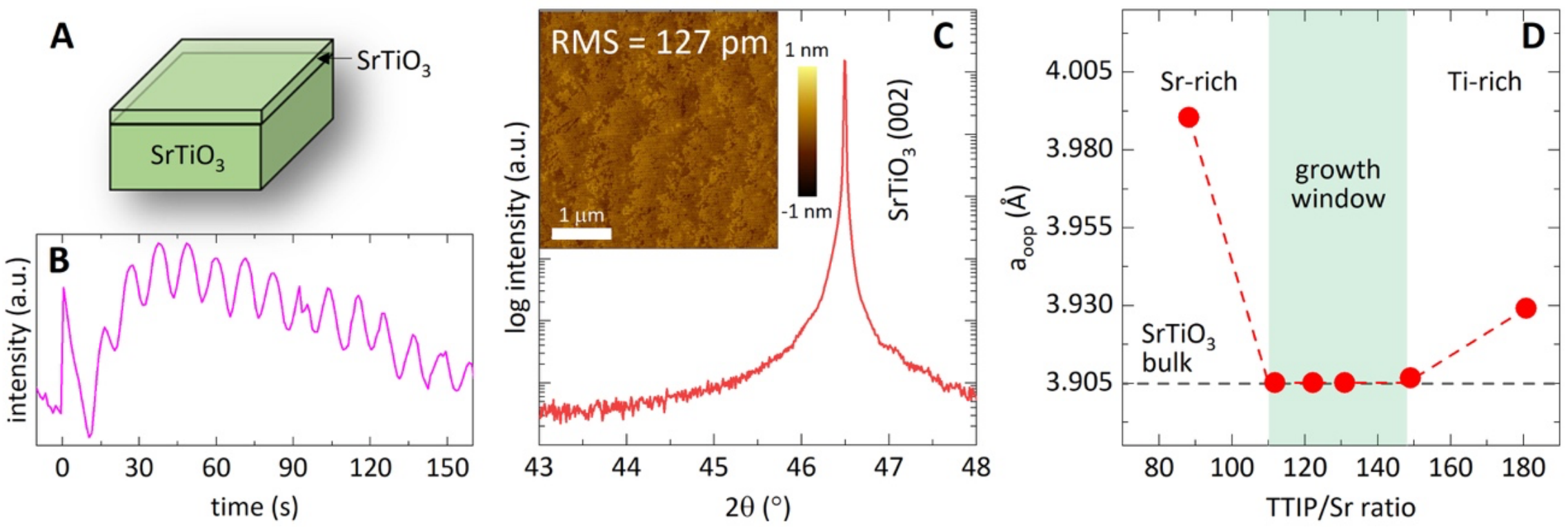

**Fig. 2. Hybrid MBE of $SrTiO_3$ without oxygen.** (**A**) Schematic of the grown film structure $SrTiO_3$/$SrTiO_3$(001). (**B**) Intensity of RHEED spots vs time during growth. (**C**) High-resolution X-ray diffraction (HRXRD) $2\theta$-$\omega$ coupled scans and AFM (inset) of the resulting film. (**D**) The lattice parameter as a function of TTIP:Sr beam equivalent pressure (BEP) ratio during growth showing the presence of a MBE growth window.

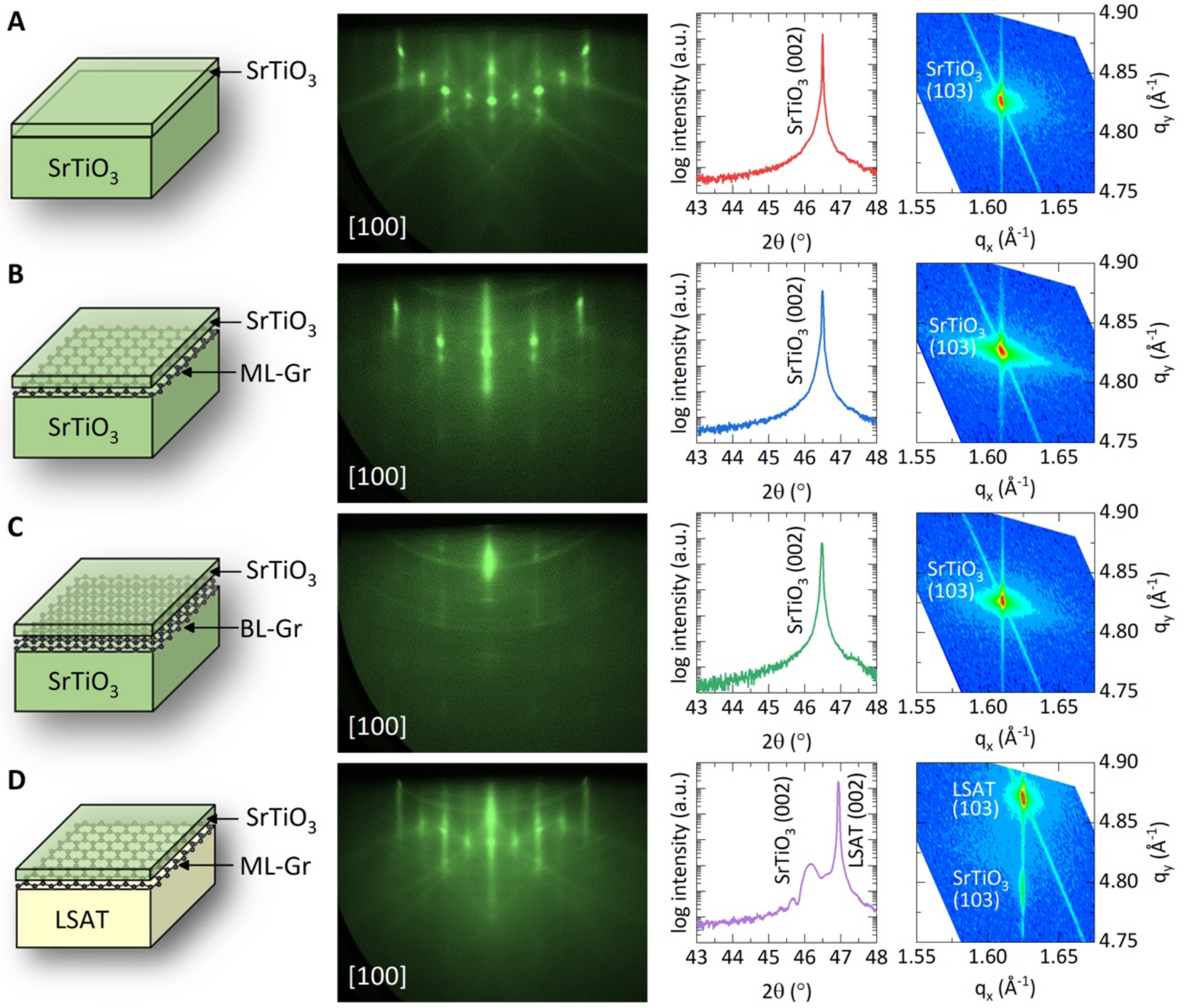

**Fig. 3. Demonstration of epitaxy for perovskites on graphene using hybrid MBE.** Sample schematic, reflection high-energy electron diffraction (RHEED), high-resolution X-ray diffraction $2\theta$-$\omega$ coupled scans, and reciprocal space maps (RSMs) of (**A**) $SrTiO_3$/$SrTiO_3$(001), (**B**) $SrTiO_3$/ML-Gr/$SrTiO_3$(001), (**C**) $SrTiO_3$/BL-Gr/$SrTiO_3$(001), and (**D**) $SrTiO_3$/ML-Gr/LSAT(001).

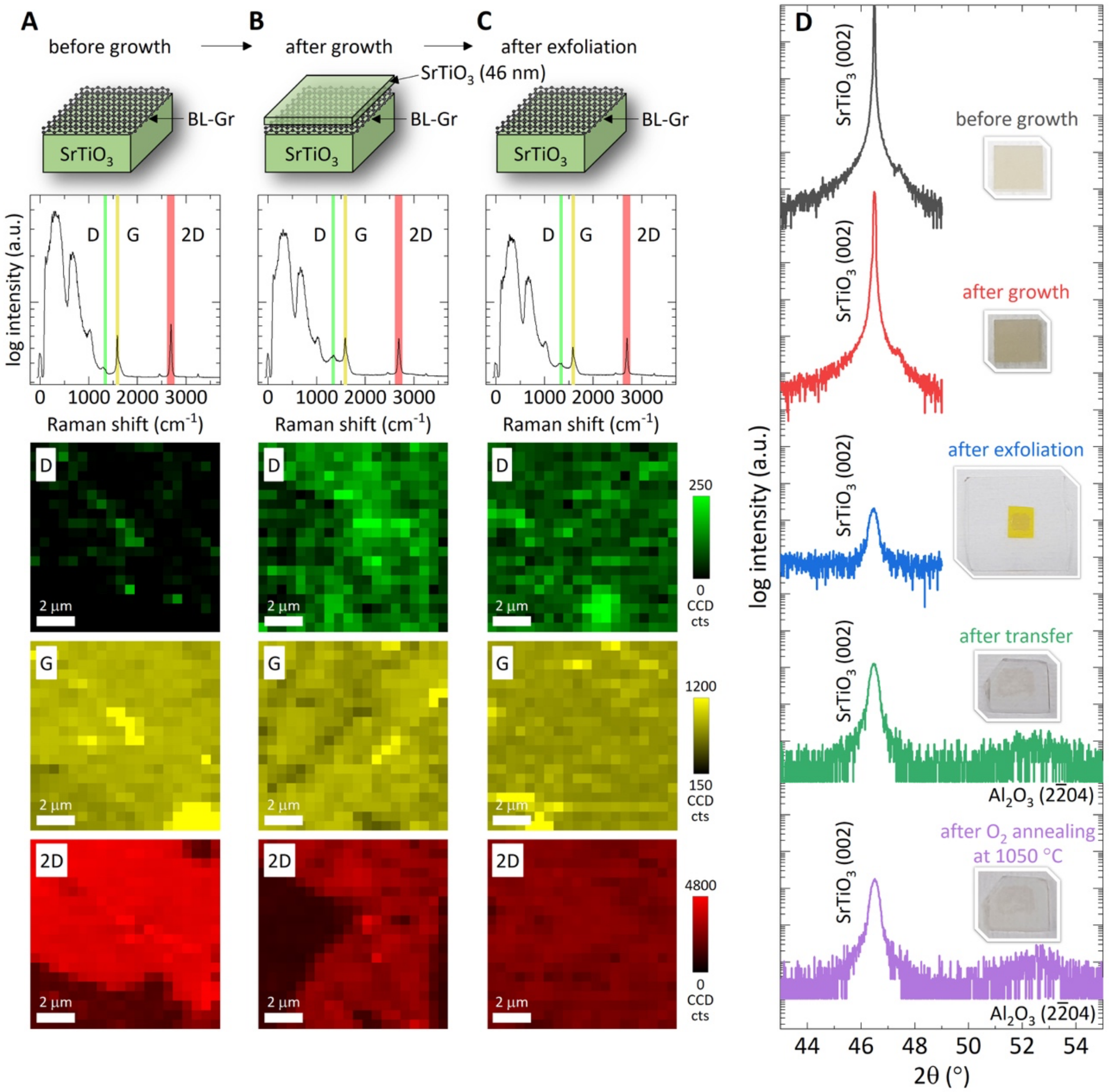

**Fig. 4. Demonstration of hybrid MBE-grown film exfoliation.** (**A**-**C**) Confocal Raman spectroscopy and microscopy of BL-Gr/$SrTiO_3$(001) before growth (**A**), the resulting $SrTiO_3$/BL-Gr/$SrTiO_3$(001) after growth (**B**), and the restored BL-Gr/$SrTiO_3$(001) via exfoliating the grown film (**C**). Each Raman micrograph shows the integrated intensity from one graphene peak scanned over the surface of the sample. (**D**) High-resolution X-ray diffraction (HR-XRD) $2\theta$-$\omega$ coupled scans of the sample before growth, and after growth, exfoliation, and then transfer to an r-plane $Al_2O_3$ substrate.

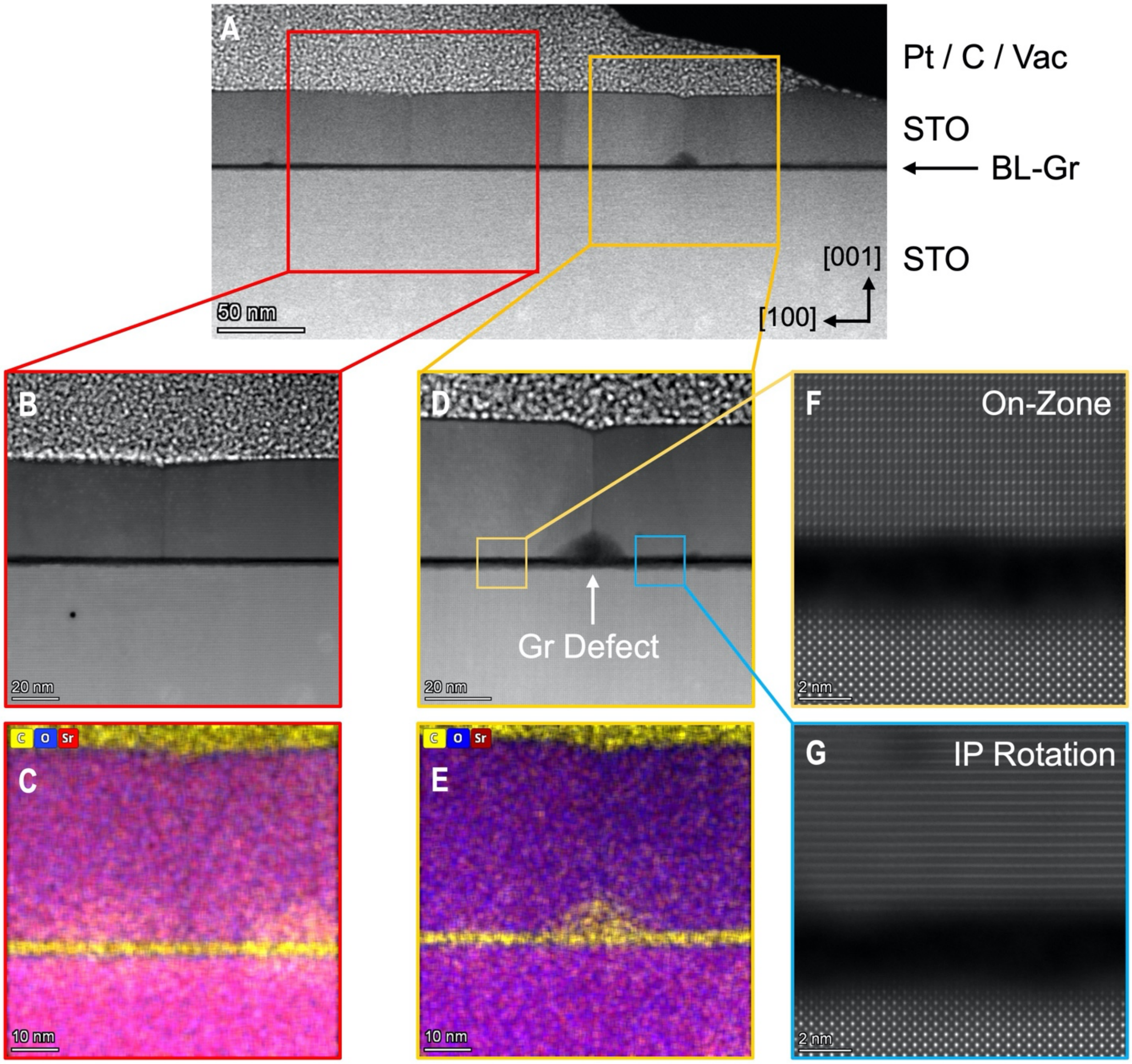

**Fig. 5. STEM characterization of pre-transfer $SrTiO_3$.** (A) Overview of a large film region containing pristine and defective graphene. STEM-HAADF images and STEM-EDS composition maps for the pristine (B-C) and defective (D-E) graphene regions, respectively. (F-G) Detail of the pristine and defective regions, showing that the former is on-zone, while the latter is rotated slightly in-plane near the defect. IP refers to in-plane.

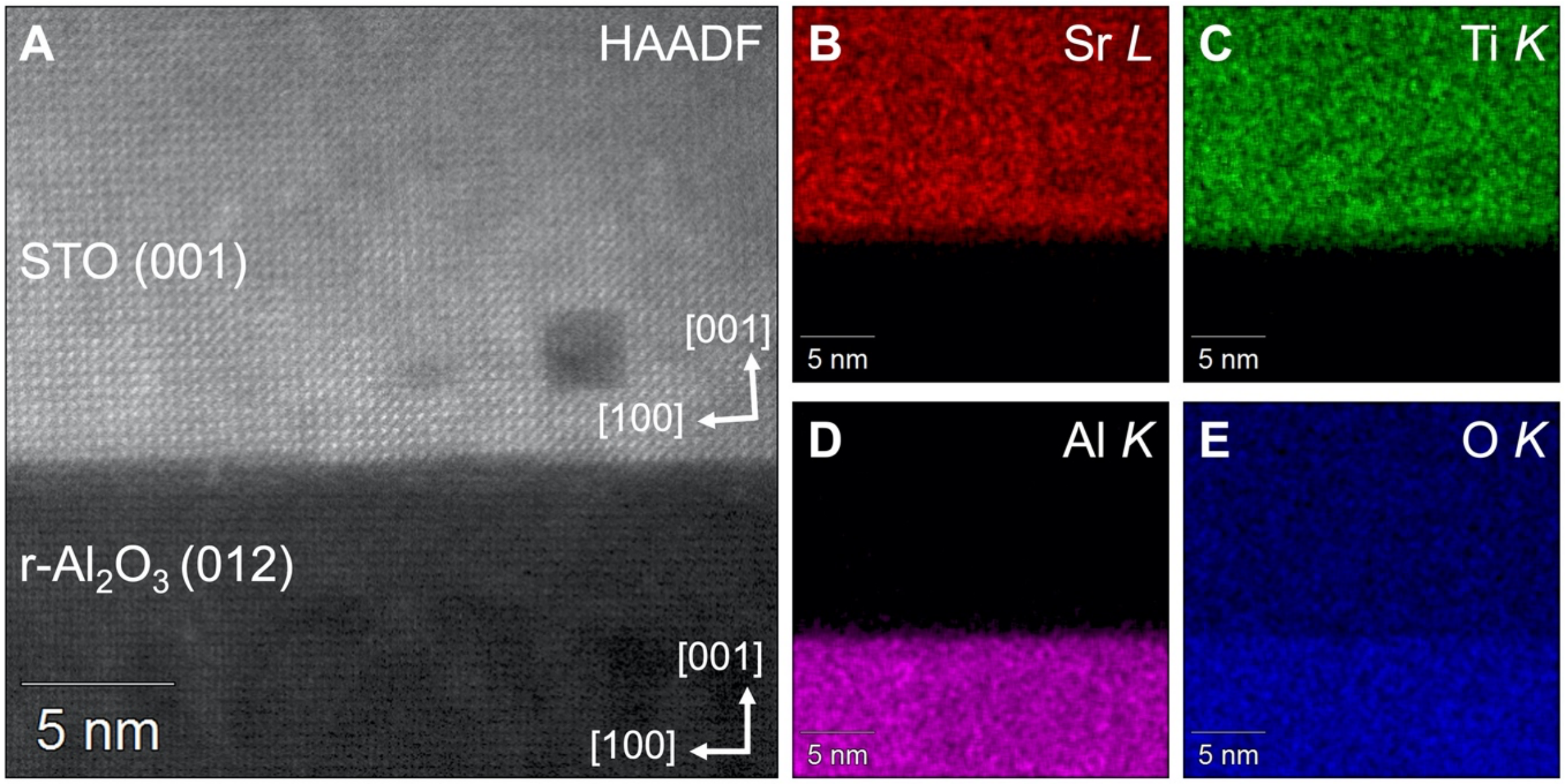

**Fig. 6. STEM characterization of transferred $SrTiO_3$ on r-$Al_2O_3$.** Scanning transmission electron microscopy of a $SrTiO_3$ epitaxial nanomembrane transferred to a foreign r-plane $Al_2O_3$ substrate. The greyscale image shows a STEM-HAADF image whereas the colored images show the STEM-EDS elemental maps of the same region. The dark square region indicates a potential void in the film. Lattice vectors are given relative to the $R\bar{3}c$ (r-$Al_2O_3$) and $Pm\bar{3}m$ (STO) space groups, respectively.

# Supplementary Information

## Free-Standing Epitaxial $SrTiO_3$ Nanomembranes via Remote Epitaxy using Hybrid Molecular Beam Epitaxy

Hyojin Yoon[1, †], Tristan K. Truttmann[1, †], Fengdeng Liu[1], Bethany E. Matthews[2], Sooho Choo[1], Qun Su[3], Vivek Saraswat[4], Sebastian Manzo[4], Michael S. Arnold[4], Mark E. Bowden[5], Jason K. Kawasaki[4], Steven J. Koester[3], Steven R. Spurgeon[2,6], Scott A. Chambers[7] and Bharat Jalan[1,*]

[1]Department of Chemical Engineering and Materials Science, University of Minnesota – Twin Cities, Minneapolis, MN 55455, USA

[2]Energy and Environment Directorate, Pacific Northwest National Laboratory, Richland, Washington 99352, USA

[3]Department of Electrical and Computer Engineering, University of Minnesota – Twin Cities, Minneapolis, MN 55455, USA

[4]Department of Materials Science and Engineering, University of Wisconsin – Madison, Madison, WI 53706, USA

[5]Environmental Molecular Sciences Laboratory, Pacific Northwest National Laboratory, Richland, Washington 99352, USA

[6]Department of Physics, University of Washington, Seattle, Washington 98195, USA

[7]Physical and Computational Sciences Directorate, Pacific Northwest National Laboratory, Richland, Washington 99352, USA

[†] HY and TKT are equally contributing authors.

* All correspondence should be addressed to BJ (bjalan@umn.edu).

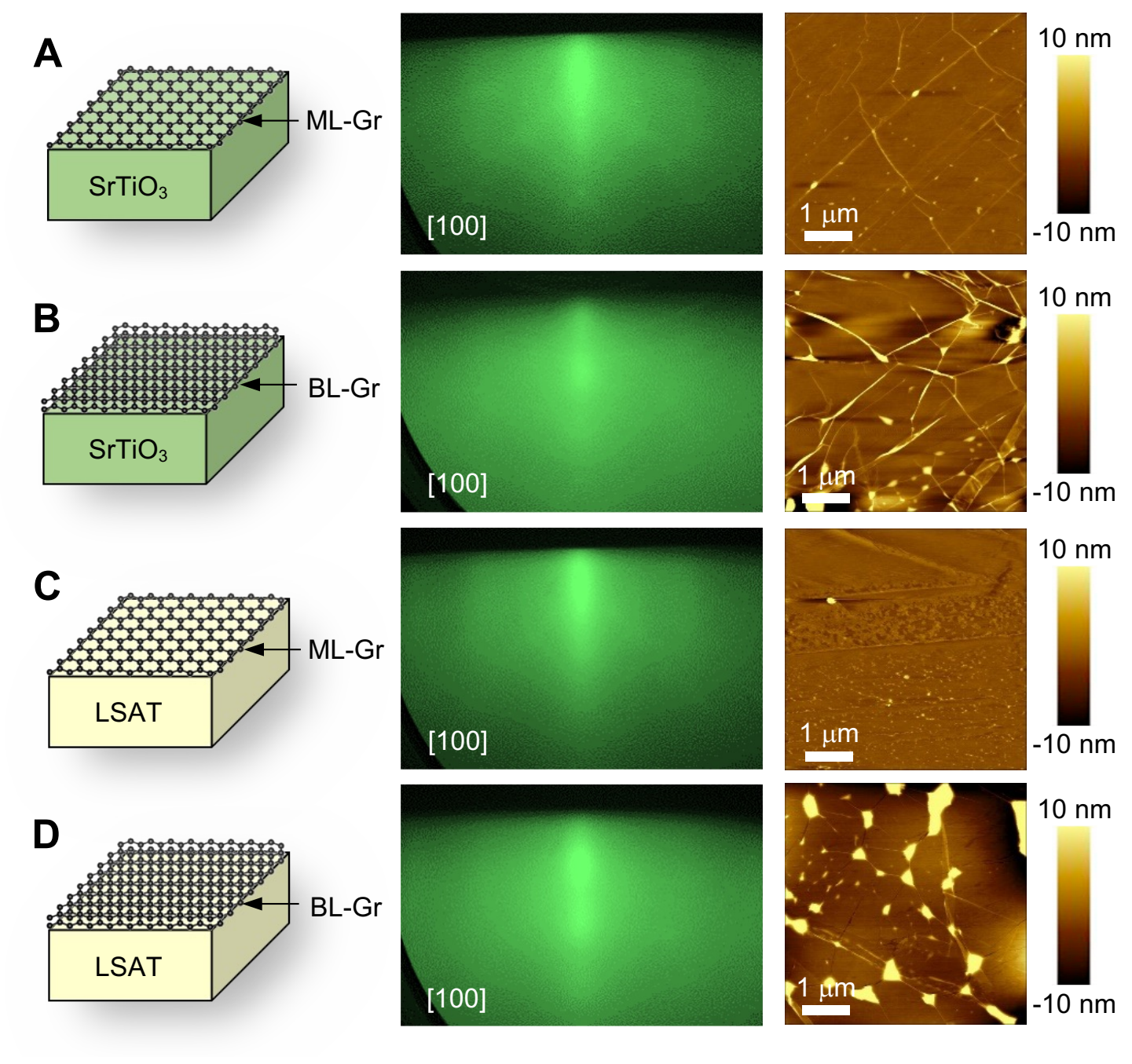

**Fig. S1. Surface structure of wet-transferred graphene.** Structure schematics, reflection high-energy electron diffraction (RHEED), and atomic force microscopy (AFM) of ML-Gr/$SrTiO_3$(001) (**A**), BL-Gr/$SrTiO_3$(001) (**B**), ML-Gr/LSAT(001) (**C**), BL-Gr/LSAT(001) (**D**). The RHEED shows the characteristic pattern of polycrystalline graphene and the AFM shows residual PMMA from the graphene wet-transfer process.

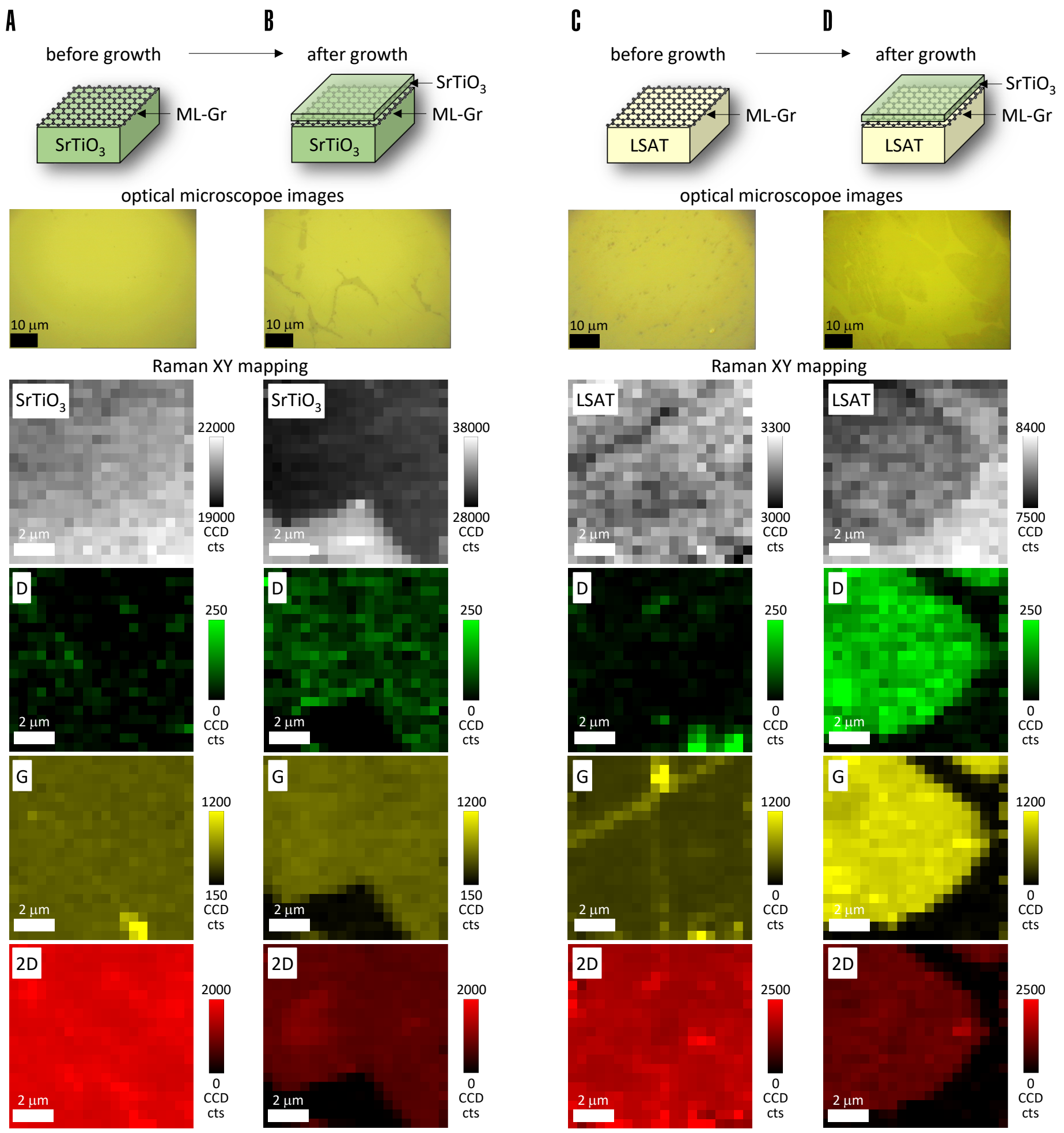

**Fig. S2. Demonstration of cracked films grown on monolayer graphene (ML-Gr).** Optical micrographs and confocal Raman micrographs of ML-Gr/$SrTiO_3$(001) before growth (**A**) and the resulting $SrTiO_3$/ML-Gr/$SrTiO_3$(001) after growth (**B**) as well as ML-Gr/LSAT(001) before growth (**C**) and the resulting $SrTiO_3$/ML-Gr/LSAT(001) after growth (**D**). All confocal Raman micrographs are rastered over the sample surface. The comparison of the micrographs shows that both the oxide films and the graphene crack during growth.

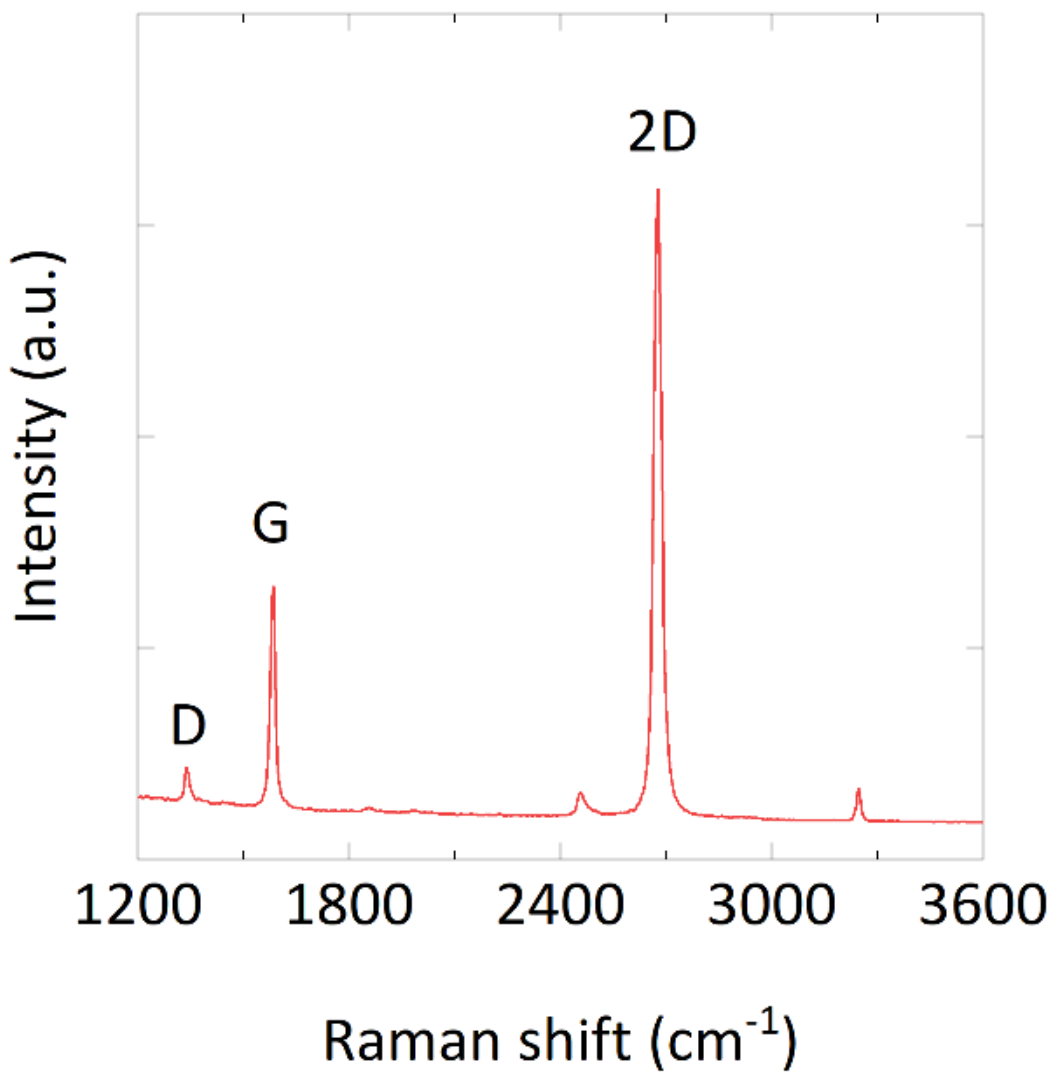

**Fig. S3**. **Determination of graphene thickness.** Raman spectroscopy of graphene transferred onto $SiO_2$/Si substrate. The intensity ratio, $I_{2D}/I_G$, is ~ 2.46, suggesting nearly monolayer thickness of the transferred graphene layer(*1*). Note that the STO substrate was not used for this analysis because of the overlap of STO and graphene signal at G peak position.

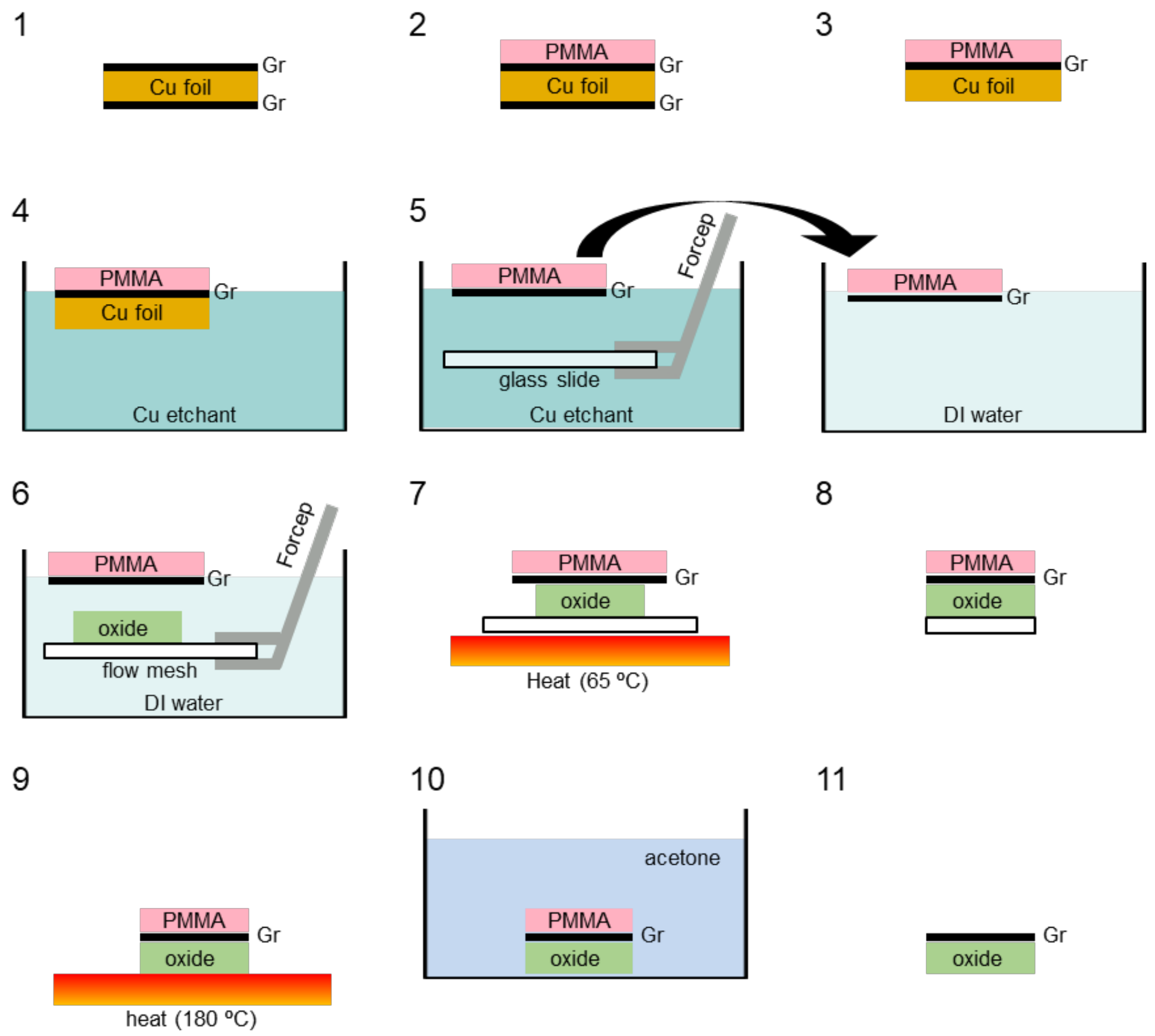

**Fig. S4. Summary of the graphene growth and wet transfer process.** (1) Graphene is synthesized on the surface of polycrystalline copper foil using chemical vapor deposition. (2) PMMA 950 C4 is spin-coated on top of the copper foil and baked on a hot plot at 180 ºC for 15 minutes. (3) The bottom-side graphene is removed using oxygen plasma. (4) The copper sheet is suspended on top of a copper etchant solution using surface tension until all copper is etched away. (5) The remaining graphene is extracted from the copper etchant using a glass slide and suspended on the surface of deionized (DI) water for cleaning. (6) The cleaned graphene is scooped onto the oxide substrate using a flow-mesh. (7) The graphene-scooped oxide substrate is heated to 65 ºC until dry. (8) The excess graphene is removed by cutting with a razor. (9) After removing the flow mesh, the sample is further baked at 180 ºC for 15 minutes. (10) The baked sample is placed in acetone until all the PMMA is dissolved away. (11) The sample is finally rinsed in acetone and isopropyl alcohol (IPA). If bilayer graphene is desired, the process is repeated once more.

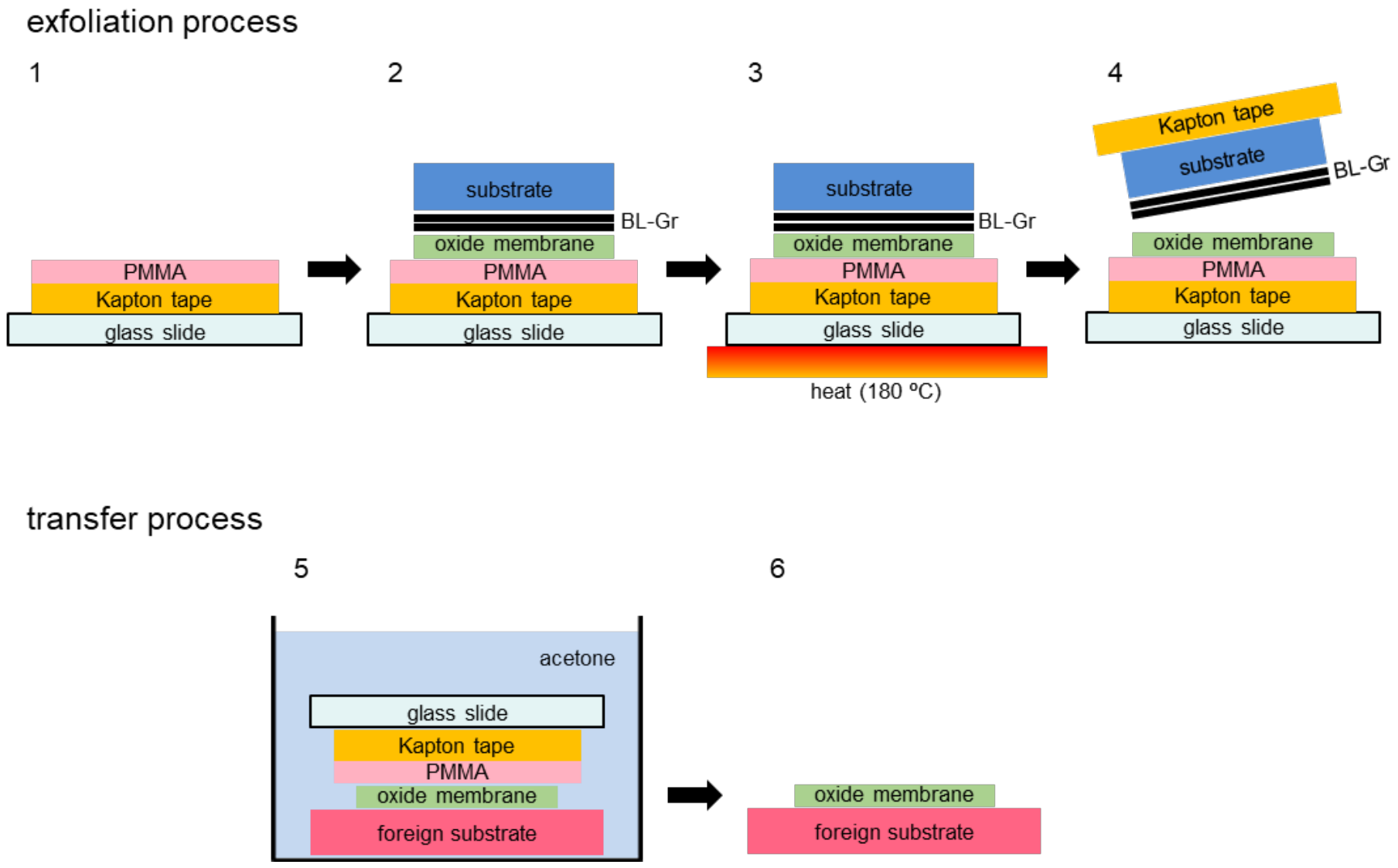

**Fig. S5. Summary of the perovskite film exfoliation and transfer process.** (1) PMMA 950 C4 is spin-coated onto Kapton tape adhered to a glass microscope slide for support. (2) The oxide film grown on bilayer graphene is gently placed on top, and air bubbles between the oxide film and the PMMA are removed by pressing on the substrate with a cotton swab. (3) The sample is heated to 180 ºC for 15 minutes to evaporate the chlorobenzene solvent. (4) The oxide film is exfoliated from the substrate by attaching Kapton tape to the back of the substrate and pulling on the tape. (5) The exfoliated oxide film is placed on the foreign substrate and submerged in acetone until all the PMMA has dissolved away. (6) After the tape and glass slide are withdrawn, the remaining sample is rinsed in acetone and IPA.